\documentclass[graybox]{svmult}

\usepackage{mathptmx}       % selects Times Roman as basic font
\usepackage{helvet}         % selects Helvetica as sans-serif font
\usepackage{courier}        % selects Courier as typewriter font
\usepackage{type1cm}        % activate if the above 3 fonts are
\usepackage{makeidx}         % allows index generation
\usepackage{graphicx}        % standard LaTeX graphics tool
\usepackage{multicol}        % used for the two-column index
\usepackage[bottom]{footmisc}% places footnotes at page bottom

\makeindex             % used for the subject index
\begin{document}

\title*{Jet tomography in heavy-ion collisions - Challenges, Results, and Open Problems}
\titlerunning{Jet tomography in heavy-ion collisions} 
\author{Barbara Betz}
\institute{Barbara Betz \at Institut f\"ur Theoretische Physik, Goethe Universit\"at, Frankfurt am Main, Germany, 
\email{betz@th.physik.uni-frankfurt.de}}
%\and Name of Second Author \at Name, Address of Institute \email{name@email.address}
\maketitle

\abstract{Over the past 30 years, jet observables have proven to provide important information about the
quark-gluon plasma created in heavy-ion collisions. I review the challenges, results, and
open problems of jet physics in heavy-ion collisions, discussing the main ideas as well as some
most recent results focussing on two major jet observables, the nuclear modification factor and the
high-$p_T$ elliptic flow.}

\section{Jets in heavy-ion collisions}
\label{sec:1}
Relativistic high-energy heavy-ion collisions offer the unique possibility to study matter 
experimentally under extreme conditions of high temperature and densities in the laboratory. One
of the main challenges is to probe the quark-gluon plasma (QGP) created in such heavy-ion 
collisions. One set of observables is based on {\it jets}, sprays of particles that
are produced back-to-back due to the conservation of energy and momentum. Those jets propagate through 
the dense matter formed while depositing energy. As this jet-energy loss
inevitably leads to an attenuation of the jet \cite{bjorken,Gyulassy:1990ye,Wang:1991xy,Zakharov:1996fv,Baier:1996kr} 
this concept is referred to as {\bf jet quenching}.

The breakthrough of studying jets in heavy-ion collisions came with the start of the 
Relativistic Heavy Ion Collider (RHIC) in 2000 \cite{whitebrahms,whitephenix,whitephobos,whitestar}. 
By studying the azimuthal distribution of the back-to-back jets in Au+Au collisions at RHIC, it 
could be shown that this part of the jet which propagates through the hot and dense matter
is suppressed (or {\bf quenched}) as compared to measurements in proton+proton (p+p) or 
deuteron+gold (d+Au) collisions \cite{Adams:2003im,Adare:2008ae}. 
This result is considered as a clear signal that at RHIC energies the hot and dense QGP medium 
is only created in heavy-ion collisions. 

Over many years, the actual evolution of the jet, the creation of shock waves and possible Mach cones
\cite{Baumgardt:1975qv,Hofmann:1976dy,Betz:2010qh}, have been discussed extensively in literature. 
However, in the following I will focus on {\bf jet tomography}, an approach pursuing the concept of jet quenching:

By studying jet quenching, one should be able to characterize some properties of the medium
created. This idea is used e.g.\ in medicine by x-ray tomography where a beam of particles
traverses a medium (e.g.\ the human body). This beam is deflected and/or absorbed and its remnants are
measured in a detector. Finally, this measurement leads to an image of the interior of the human's body.
Likewise, one aims to getting an image of the interior of a heavy-ion collision by performing
jet tomography.

The basic idea of jet quenching and jet tomography has been applied in heavy-ion collisions since the 1990's. 
On the theory side, it has lead to various jet-quenching models: GLV, DGLV, WHDA, AMY, ASW, ... 
\cite{GLV,Djordjevic:2003zk,Wicks:2007am,Baier:1996sk,Wiedemann:2000za,Qiu:1990xa,Arnold:2001ba}.

\subsection{Major jet observables}
Jet quenching is predominantly quantified by the nuclear modification factor ($R_{AA}$) which is the ratio 
of the number of particles created in a nucleus+nucleus (A+A) collision scaled to the number
of particles created in a p+p collision and the number of collisions $N_{\rm coll}$:

\begin{equation}
R_{AA}(p_T) = \frac{dN_{AA}/dp_T}{N_{\rm coll}dN_{pp}/dp_T}
\end{equation}

Usually, this ratio is given as a function of the transverse momentum $p_T$. 
If a heavy-ion collision was a pure superposition of a p+p collisions then $R_{AA}=1$.
However, if there is jet quenching then $R_{AA}<1$.

One of the major results obtained at RHIC \cite{Adams:2003kv,Adler:2006bv} 
was to show that the measured nuclear modification factor for pions, the 
predominant species of particles measured in Au+Au collisions, is $R_{AA}\sim 0.2$
which is significantly below 1.

With this measurement, the predicted jet suppression \cite{bjorken,Gyulassy:1990ye,Wang:1991xy,Vitev:2002pf} 
was first observed at RHIC and it is considered as a signal for the creation of 
an opaque matter, the quark-gluon plasma \cite{Gyulassy:2004zy,Shuryak:2003xe}.

A second major jet observable, the high-$p_T$ elliptic flow ($v_2$), is based on a characteristic
observable of the background medium, the elliptic flow $v_2$. Most A+A collisions show
an offset. If the particles in the overlap region, where the hot and dense QGP medium is formed, 
interact then gradients will lead to a preferred emission. 

By comparison to hydrodynamic simulations \cite{Romatschke:2007mq}, it was 
shown that the background medium shows a preferred direction resulting an asymmetry which is 
quantified by the 2nd Fourier coefficient of the angular distribution, 
the elliptic flow $v_2$:

\begin{equation}
\frac{dN}{d\phi} = \frac{N}{2\pi}
\left[
1+2\sum\limits_{n=1}^{\infty} v_n \cos(n\phi)
\right].
\end{equation}

Jets created in the overlap region that interact with the medium will certainly be affected
by the preferred emission of the background medium, resulting in a preferred emission of the
high-$p_T$ particles (jets). Thus, even though the underlying physics leading to this preferred
emission is different, this observed preferred emission is referred to as {\it high-$p_T$ elliptic flow}.

\section{Jet tomography -- a challenge in heavy-ion collisions}

Usually the jet-energy loss in heavy-ion collisions is considered to
be very similar to tomography of X-rays routinely used in medicine. However, both procedures
are indeed quite different. 

In contrast to an ideal tomography (for simplicity one might think of an x-ray tomography 
mentioned above), a heavy-ion collision misses \cite{MG}
\begin{enumerate}
\item a controlled flow of penetrating particles,
\item an established dynamical theory of the energy loss,
\item and a non-moving, non-fluctuating background medium.
\end{enumerate}

Of course, this does {\it not} imply that jet tomography cannot be done in heavy-ion collisions
but it indicates that conclusions might not be as straightforward as they seem.

\section{Immediate consequences from the first results at the LHC}

Before the Large Hadron Collider (LHC) was turned on in 2010, one of main questions discussed was if the correct description
of a jet-energy loss in heavy-ion collisions is done by using perturbative QCD (pQCD) or by
applying the Anti-de-Sitter/Conformal Field Theory (AdS/CFT) correspondence
\cite{GLV,Djordjevic:2003zk,Wicks:2007am,Baier:1996sk,Wiedemann:2000za,Qiu:1990xa,Arnold:2001ba,Maldacena:1997re,ches1,ches2,ches3,Gyulassy:2011zz}.

Pre-LHC runs performed at RHIC \cite{Adare:2010sp} indicated that the measured nuclear modification factor 
and the high-$p_T$ elliptic flow can only be described {\bf simultaneously} if a squared
path-length dependence, $dE/dx=dE/d\tau\sim\tau^2$, is considered. This squared path-length
dependence points to an AdS/CFT-like energy loss while a pQCD-like jet-energy loss
is assumed to have a linear path-length dependence, $dE/dx=dE/d\tau\sim\tau$.

Right after the start of the LHC, a remarkable result was obtained for the nuclear
modification factor. In contrast to early pQCD-based predictions \cite{Vitev:2002pf},
the $R_{AA}$ showed an unexpected similarity for measurements at RHIC an LHC 
in the region $10<p_T<20$~GeV. 

These measurements indicate that there is a {\it surprising QGP-transparency} 
\cite{Horowitz:2011gd,Betz:2011tu,Pal:2012gf} for LHC energies. It suggests 
that the jet-medium coupling at LHC energies is smaller than at RHIC energies 
which points to a running-coupling effect consistent with pQCD
but not with AdS/CFT.

Besides this, {\it conformal} AdS/CFT energy loss was expected to yield a flat
$R_{AA}$ \cite{Horowitz:2007ui}. However, the distinct slope of the pQCD prediction
for the nuclear modification factor at LHC energies given in Ref.\ \cite{Vitev:2002pf}
was shown to be correct, resulting in the question if (conformal) AdS/CFT was ruled out 
by the first data of the LHC and how to resolve the puzzle connected to the
pre-LHC runs.

\section{The BBMG model}

To investigate the measured jet-energy loss at RHIC and LHC energies, 
we developed a generic jet-energy loss model (for convenience referred to as BBMG model) 
over the past few years \cite{Betz:2011tu,Betz:2012qq,Betz:2014cza}. This model is
based on the following ansatz of the jet-energy loss:
\begin{eqnarray}
\frac{dE}{d\tau}= 
-\kappa\,  E^a \, \tau^{z} \, T^{c=(2+z-a)} \, \zeta_q \, \Gamma_f ,
\label{Eq1}
\end{eqnarray}
with the jet-medium coupling $\kappa$, jet energy $E$, the path-length $\tau$, and the temperature density of 
the background medium $T$. 

Jet-energy loss fluctuations are included via the distribution
\begin{equation}
f_q(\zeta_q)= \frac{(1 + q)}{(q+2)^{1+q}} (q + 2- \zeta_q)^q 
\end{equation} 
which allows for an easy interpolation between non-fluctuating ($\zeta_{q=-1}=1$) distributions 
and those ones increasingly skewed towards small $\zeta_{q>-1} < 1$
\cite{Betz:2014cza}.

The background flow fields are incorporated via the flow factor 
\begin{equation}
\Gamma_f=\gamma_f [1 - v_f \cos(\phi_{\rm jet} - \phi_{\rm flow})]
\end{equation}
\cite{Liu:2006he,Baier:2006pt,Renk:2005ta,Armesto:2004vz} with the background flow velocities $v_f$ and the $\gamma$-factor
\begin{equation}
\gamma_f = 1/\sqrt{1-v_f^2}\; .
\end{equation}
$\phi_{\rm jet}$ is the jet angle w.r.t.\ the reaction plane and 
$\phi_{\rm flow}=\phi_{\rm flow}(\vec{x},t)$ is the corresponding local azimuthal
angle of the background flow fields.

Even though this model is quite simple and not based on first-principles calculations,
it has offered the possibility to explore the jet physics in high-energy heavy-ion collisions
to a surprising quantitative accuracy. Besides that, the results obtained
via the BBMG model have always been cross-checked with the CUJET model 
\cite{Xu:2014tda,Buzzatti:2012dy,Buzzatti:2012pe} which based on pQCD calculations up 
to 10 orders in opacity. 

The BBMG model interpolates between pQCD-based and AdS/CFT-inspired 
jet-energy loss algorithms with a linear and a squared path-length $z$, respectively, and has 
been coupled to state-of-the-art hydrodynamic and parton cascade background media
\cite{VISH2+1,Luzum:2008cw,BAMPS}.

By calculating the nuclear modification factor and the high-$p_T$ elliptic flow
for RHIC and LHC, the jet-medium coupling, the jet-energy dependence, the
path-length dependence, and the impact of the background have been explored
\cite{Betz:2011tu,Betz:2012qq,Betz:2014cza}. 

\begin{figure}[t]
\sidecaption[t]
\includegraphics[scale=.65]{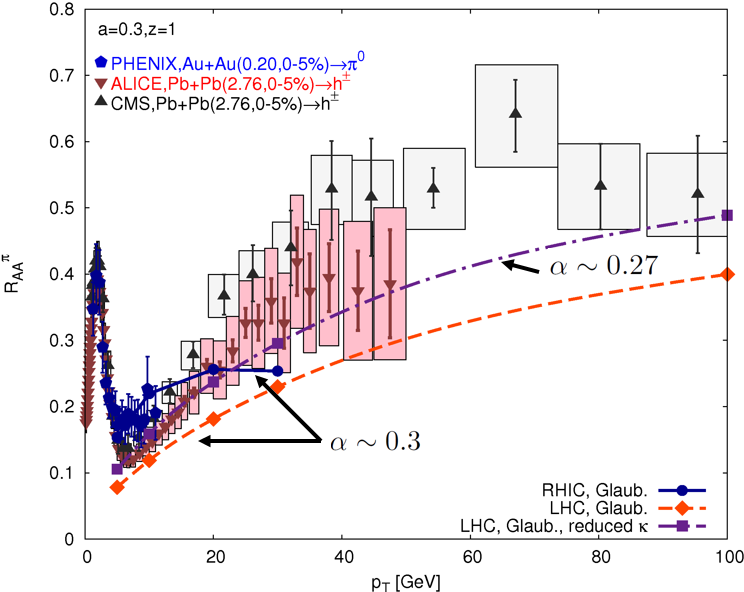}
\caption{The early results for the nuclear modification factor measured at RHIC 
and LHC \cite{PHENIX_old,Collaboration:2011hf,CMSRAA} compared to calculations
of an early version of the BBMG model given by $dE/d\tau= -\kappa\,  E^{a=0.3} \tau^{z=1} \, T^{c=2.7}$
without jet-energy loss fluctuations and background flow. The results for a coupling constant of $\alpha=0.3$ 
reproduce the measured data at RHIC energies (blue solid line) but overquench at LHC energies 
(orange dashed line). For a moderately reduced coupling of $\alpha=0.27$, the nuclear 
modification factor for pions at LHC energies (magenta dashed-dotted line) is preferred.}
\label{Fig01}       
\end{figure}

In particular, we have been able to show that a moderate reduction of the jet-medium coupling
is needed to describe the LHC nuclear modification factor at LHC energies, see
Fig.\ \ref{Fig01}. We could also prove that the rapid rise of the nuclear 
modification factor at LHC energies rules out any model with $dE/dx\sim E^{a>1/3}$.
This rapid rise can easily seen from Fig.\ \ref{Fig01}.
Please note that $a=1/3$ is the lower bound of the falling-string scenario, 
while $a=0$ is referred to a pQCD-scenario \cite{Betz:2012qq}.

By performing a detailed survey \cite{Betz:2014cza}, we demonstrated that a pQCD-based
scenario with the parameters $a=0, z=1, c=3$ in Eq.\ (\ref{Eq1}) describes
the measured nuclear modification factor and the high-$p_T$ elliptic flow
within the uncertainties of the bulk evolution if a running jet-medium coupling is considered.
Those uncertainties are given e.g.\ by the initial state and the viscosity of the background medium.

In case of a {\it conformal} AdS-scenario with a squared path-length dependence, however, the nuclear modification
factor is clearly overquenched \cite{Betz:2014cza}. The reason is that a conformal AdS-scenario is 
characterized by a fixed jet-medium coupling since a conformal theory does not 
have any additional scale which can run. Thus, we concluded \cite{Betz:2014cza}
that a {\it conformal} scenario is ruled out by the rapid rise of the measured
$R_{AA}(p_T)$. 

In contrast, a {\it non-conformal} AdS-scenario \cite{Ficnar:2011yj,Finazzo:2014cna} 
allowing for a running of the jet-medium coupling, does lead to similar results 
as the pQCD calculations \cite{Betz:2014cza}. Thus, we observed that a linear and 
a squared path-length dependence lead to similar results for the nuclear 
modification factor and the high-$p_T$ elliptic flow and does not allow for an 
any disentangling of a possible pQCD and AdS-scenario \cite{Betz:2014cza}. One 
of the main open challenges is to find a possible new observable 
which breaks this degeneracy.

\section{The high-$p_T$ $v_2$ problem}

\begin{figure}[t]
\sidecaption[t]
\hspace*{-0.2cm}
\includegraphics[scale=.52]{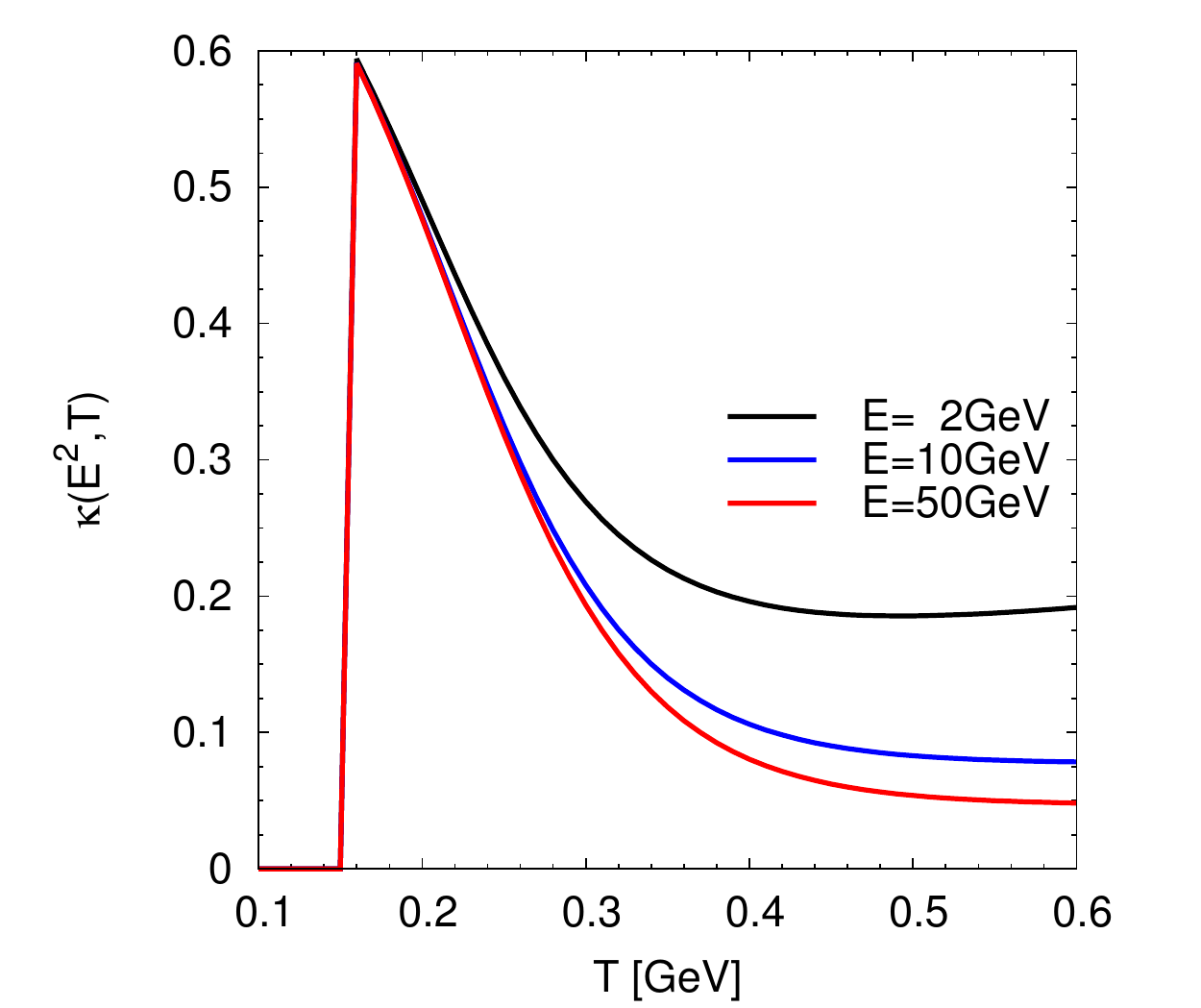}
\caption{The jet-medium coupling $\kappa(E^2,T)$, generalized in Ref.\
\cite{Xu:2014tda}, as a function of the jet energy $E$ and the medium temperature $T$.
There is no interaction below $T_c=0.16$~GeV as it is assumed that the
medium is converted into hadrons below this temperature.}
\label{Fig02}       
\end{figure}

As mentioned above, we showed in Ref.\ \cite{Betz:2014cza} that a pQCD scenario describes
the measured data within the theoretical and experimental uncertainties given.
However, our results are at the lower end of the measured
error bars. This is in line with other jet-energy loss models. While 
various different models can describe the $R_{AA}$, the
high-$p_T$ is rather challenging and up to a factor of 2
too small as compared to the data \cite{Adare:2010sp,Xu,Molnar}.

Ref.\ \cite{Xu:2014tda} suggested that a jet-medium coupling 
including non-perturbative effects around the phase 
transition at $T_c\sim160$~GeV and depending both on the energy of 
the jet and the temperature of the background medium,
$\kappa(E^2,T)$, resolves the high-$p_T$ $v_2$ problem and leads to a 
simultaneous prescription of both the $R_{AA}$ and the high-$p_T$ $v_2$. 

Since the jet-medium coupling $\kappa(E^2,T)$ has been generalized 
to an analytic form \cite{Xu:2014tda} (which is plotted in Fig.\
\ref{Fig02}), it can easily be included in the BBMG model
given by Eq.\ (\ref{Eq1}). In Ref.\ \cite{Betz:2015mlf} we showed
that the energy and temperature-dependent jet-medium coupling improves
the description of the high-$p_T$ elliptic flow drastically, independently
of a hydrodynamic or parton cascade background medium considered.

\section{SHEE -- Soft-Hard Event Engineering}

\begin{figure}[t]
\sidecaption[t]
\includegraphics[scale=.52]{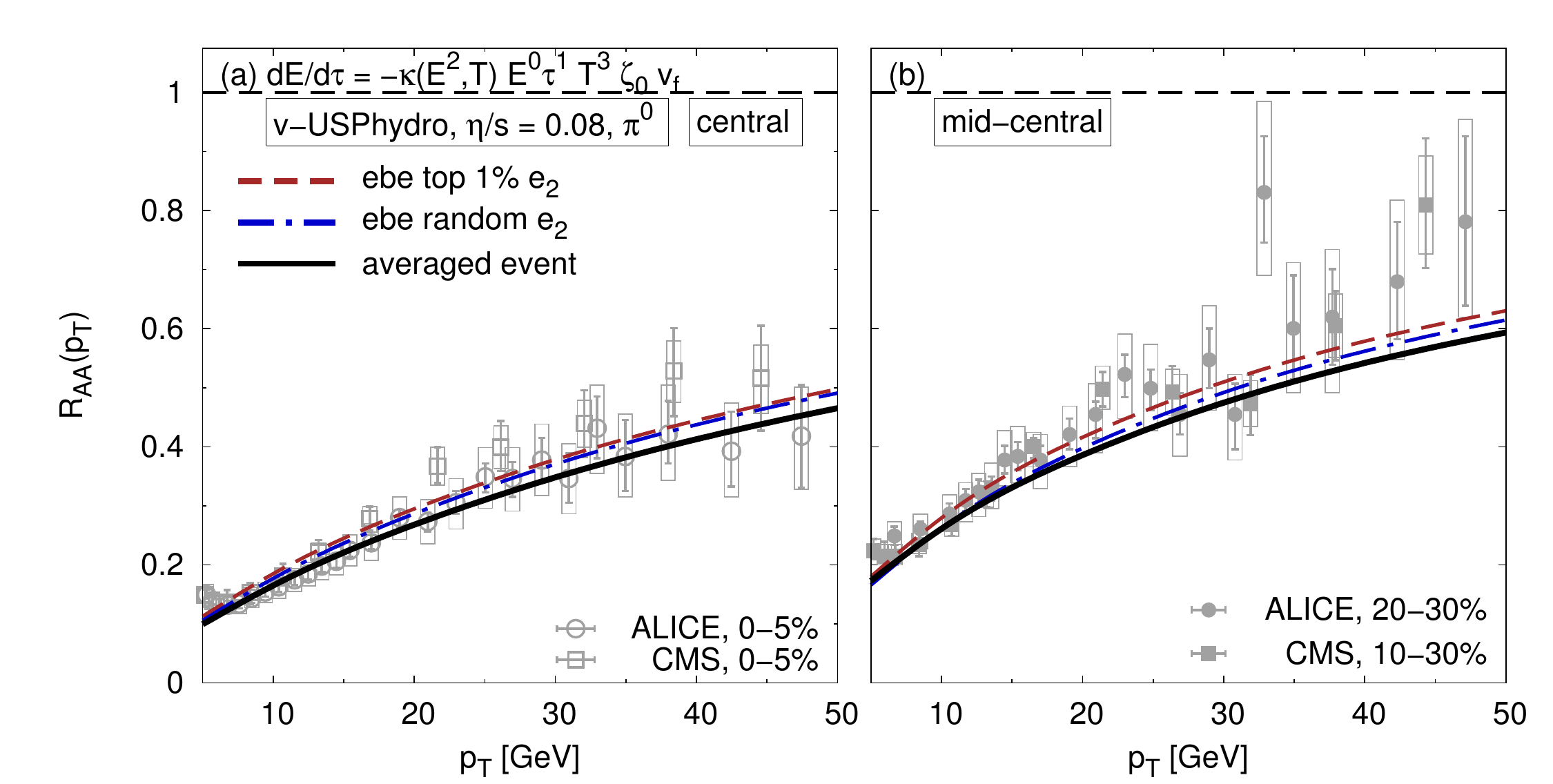}
\caption{The nuclear modification factor for central and mid-central events
measured at $\sqrt{2.75}$~TeV LHC energy \cite{data2_RAA,data3_RAA} 
compared to the three Glauber $e_2$-eccentricity selections of the centrality
classes 0-5\% (left) and 20-30\% (right).}
\label{Fig03}       
\end{figure}

To further study the impact of the background on the high-$p_T$ elliptic flow,
in particular the impact of the $e_2$-eccentricity selection (determining
the centrality of a collision) within a given centrality class, we have 
recently started to compare various selected soft (low-$p_T$ background) and
hard (high-$p_T$ jet) events \cite{BBMGvUSPhydro}.

The wide low-$p_T$ distributions measured by the ATLAS collaboration \cite{Aad:2013xma}
have proven that background models must render both the $\langle {\rm low-}p_T v_n \rangle$
and the correct fluctuations within a centrality class.

For SHEE \cite{Noronha-Hostler:2016eow}, we coupled the (hydrodynamic) v-USPhydro code \cite{Noronha-Hostler:2013gga,Noronha-Hostler:2014dqa}
to the BBMG model. 15,000 Glauber initial conditions are generated and three
different events are selected:
\begin{enumerate}
\item 150 events with random $e_2$-eccentricity,
\item 150 events with top 1\% $e_2$-eccentricity,
\item and an averaged event (smoothed profile).
\end{enumerate}

\begin{figure}[t]
\sidecaption[t]
\includegraphics[scale=.5]{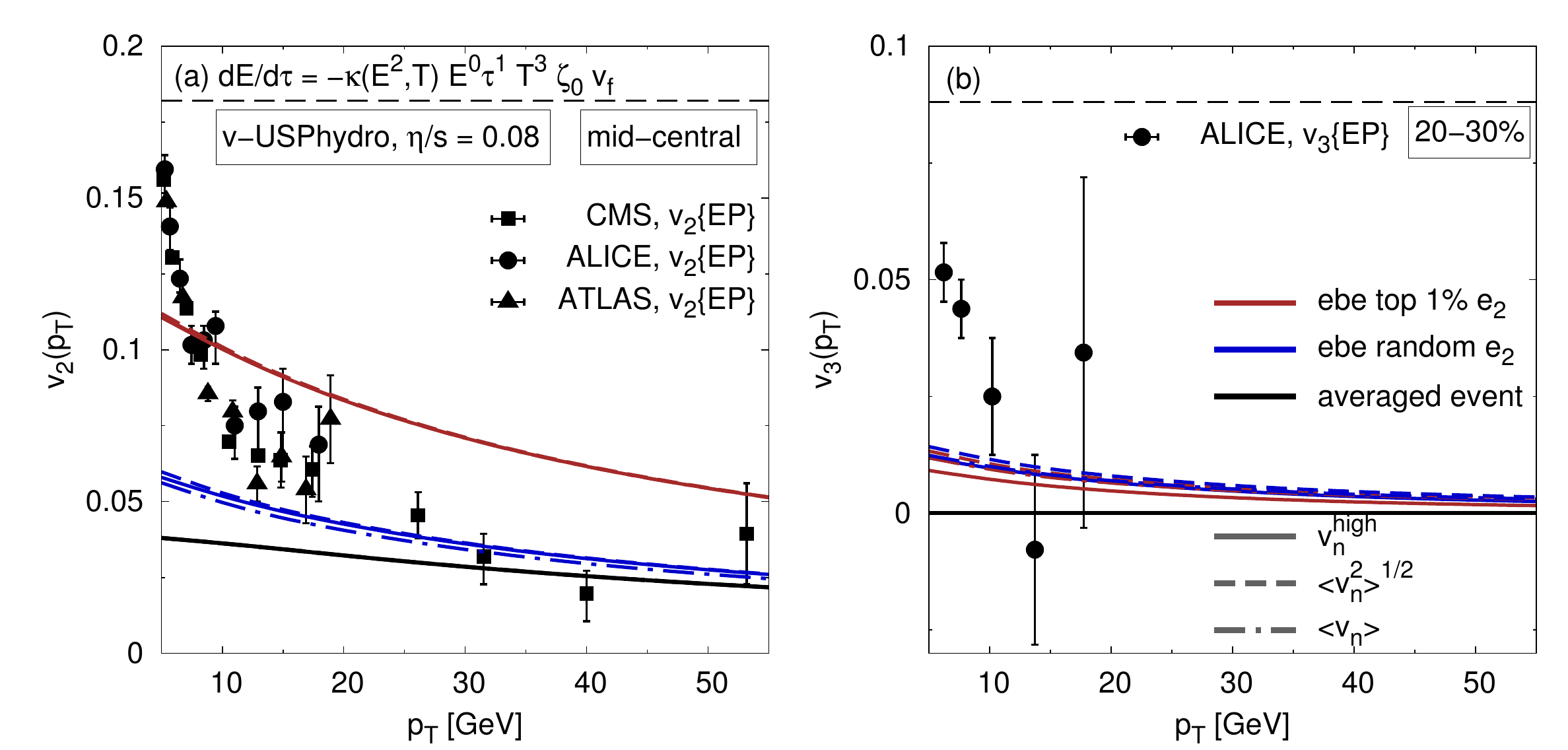}
\caption{The high-$p_T$ $v_2$ (left) and $v_3$ (right) calculated via the arithmetic mean, the 
root mean square, and Eq.\ (\ref{Eq_highvn}) for the three $e_2$-eccentricity 
selections of the centrality classes 20-30\% and compared to the
measured data at $\sqrt{2.75}$~TeV LHC energy \cite{data2_v2,data3_v2,data4}.}
\label{Fig04}  
\end{figure}

Those initial conditions are consecutively run through the v-USPhydro and BBMG 
code. The results for the nuclear modification factor and the high-$p_T$ elliptic 
flow are shown in Figs.\ \ref{Fig03} and \ref{Fig04}. Please note that the reference point
chosen for all scenarios is $R_{AA}(p_T=10\,{\rm GeV})=0.185$. In Fig.\
\ref{Fig04} we compare three different methods to determine the high-$p_T$ elliptic 
flow: the arithmetic mean $\langle v_n\rangle$, the root mean square 
$\langle v_n^2\rangle^{1/2}$, and $v_n^{\rm high}$ given by
\begin{equation}
v_n^{\rm high} = \frac{\langle v_n^{\rm low} v_n^{\rm high}(p_T) \cos[n(\psi_n^{\rm low}-\psi_n^{\rm high}(p_T))]\rangle_{\rm events}}
{\sqrt{\langle v_n^{2,\rm{low}}\rangle_{\rm events}}}
\label{Eq_highvn}
\end{equation}
which is used by experiment \cite{Aad:2015lwa,Aamodt:2011by}.

Fig.\ \ref{Fig03} shows that there is almost no difference between the event-by-event
and smoothed initial conditions for the nuclear modification factor. Thus, 
$R_{AA}$ is independent of the $e_2$-eccentricity distribution of the background
medium.

Fig.\ \ref{Fig04} demonstrates that the high-$p_T$ $v_2$ is proportional to the
low-$p_T$ $v_2$ (which is the largest for the top 1\% $e_2$ events) and 
that the width of the low-$p_T$ $v_2$-distribution
influences the high-$p_T$ $v_2$. Besides that, the event-by-event fluctuations
enhance the high-$p_T$ $v_2$, depending on the $e_2$-eccentricity selection. 
The yields for the arithmetic mean $\langle v_n\rangle$, the root mean square 
$\langle v_n^2\rangle^{1/2}$, and the $v_n^{\rm high}$ are similar. 
Fig.\ \ref{Fig04} exhibits that $e_2$ and $e_3$ are anticorrelated as for the 
low-$p_T$ bulk medium \cite{Niemi:2012aj,Retinskaya:2013gca}.

\section{Conclusions}

By reviewing the challenges, results, and open problems of jet physics in 
heavy-ion collisions, discussing the main ideas and concepts as well as 
most recent results of the nuclear modification factor and the high-$p_T$ elliptic flow,
I showed that jets are important tools to probe heavy-ion collisions. 
Unfortunately, it has not yet been possible to disentangle the underlying
theory (pQCD or AdS/CFT) with these measurements but we must find a way.
Jet physics in heavy-ion collision clearly shows that the results obtained
are influenced both by the background medium and the jet-energy loss
description.  

\section{Acknowledgements}

The author thanks C.\ Greiner, M.\ Gyulassy, J.\ Noronha-Hostler, J.\ Noronha, F.\ Senzel, 
and J.\ Xu for the fruitful collaboration as well as U.\ Heinz and C.\ Shen for 
providing their hydrodynamic field grids. This work was supported 
through the Bundesministerium f\"ur Bildung und Forschung under project number
05P2015, the Helmholtz International Centre for FAIR within the framework 
of the LOEWE program (Landesoffensive zur Entwicklung 
Wissenschaftlich-\"Okonomischer Exzellenz) launched by the State of Hesse.
The author thanks the organizers of the Symposium on "New Horizons in Fundamental Physics"
for an inspriring workshop celebrating Walter Greiner's 80th birthday.

%%%%%%%%%%%%%%%%%%%%%%%% referenc.tex %%%%%%%%%%%%%%%%%%%%%%%%%%%%%%
% sample references
% %
% Use this file as a template for your own input.
%
%%%%%%%%%%%%%%%%%%%%%%%% Springer-Verlag %%%%%%%%%%%%%%%%%%%%%%%%%%
%
% BibTeX users please use
% \bibliographystyle{}
% \bibliography{}

\begin{thebibliography}{99.}%
% Use the following syntax and markup for your references if 
% the subject of your book is from the field 
% "Mathematics, Physics, Statistics, Computer Science"
%

% Journal article
\bibitem{bjorken}
  J.~D.~Bjorken, Phys.\ Rev.\ D {\bf 27}, 140 (1983).

\bibitem{Gyulassy:1990ye}
  M.~Gyulassy and M.~Plumer, Phys.\ Lett.\  B {\bf 243}, 432 (1990).  
   
\bibitem{Wang:1991xy}
  X.~N.~Wang and M.~Gyulassy, Phys.\ Rev.\ Lett.\  {\bf 68}, 1480 (1992).  

\bibitem{Zakharov:1996fv}
  B.~G.~Zakharov, JETP Lett.\  {\bf 63}, 952 (1996).

\bibitem{Baier:1996kr}
  R.~Baier, Y.~L.~Dokshitzer, A.~H.~Mueller, S.~Peigne and D.~Schiff, Nucl.\ Phys.\  B {\bf 483}, 291 (1997).

\bibitem{whitebrahms}
 I.~Arsene {\it et al.}  [BRAHMS Collaboration], Nucl.\ Phys.\ A {\bf 757}, 1 (2005).

\bibitem{whitephenix}
 K.~Adcox {\it et al.}  [PHENIX Collaboration], Nucl.\ Phys.\ A {\bf 757}, 184 (2005).

\bibitem{whitephobos}
  B.~B.~Back {\it et al.}, Nucl.\ Phys.\ A {\bf 757}, 28 (2005).

\bibitem{whitestar}
  J.~Adams {\it et al.}  [STAR Collaboration], Nucl.\ Phys.\ A {\bf 757}, 102 (2005).

\bibitem{Adams:2003im} 
  J.~Adams {\it et al.} [STAR Collaboration], Phys.\ Rev.\ Lett.\  {\bf 91}, 072304 (2003).

\bibitem{Adare:2008ae} 
  A.~Adare {\it et al.} [PHENIX Collaboration], Phys.\ Rev.\ C {\bf 78}, 014901 (2008).

\bibitem{Baumgardt:1975qv} 
  H.~G.~Baumgardt, J.~U.~Schott, Y.~Sakamoto, E.~Schopper, H.~Stoecker, J.~Hofmann, W.~Scheid and W.~Greiner, 
  Z.\ Phys.\ A {\bf 273}, 359 (1975).

\bibitem{Hofmann:1976dy} 
  J.~Hofmann, H.~Stoecker, U.~W.~Heinz, W.~Scheid and W.~Greiner, 
  Phys.\ Rev.\ Lett.\  {\bf 36}, 88 (1976).

\bibitem{Betz:2010qh} 
  B.~Betz, J.~Noronha, G.~Torrieri, M.~Gyulassy and D.~H.~Rischke,
  Phys.\ Rev.\ Lett.\  {\bf 105}, 222301 (2010).

\bibitem{GLV}
  M.~Gyulassy, P.~Levai and I.~Vitev, Phys.\ Rev.\ Lett.\  {\bf 85}, 5535 (2000).

\bibitem{Djordjevic:2003zk}
  M.~Djordjevic and M.~Gyulassy, Nucl.\ Phys.\  A {\bf 733}, 265 (2004).

\bibitem{Wicks:2007am}
  S.~Wicks, W.~Horowitz, M.~Djordjevic and M.~Gyulassy, Nucl.\ Phys.\  A {\bf 783}, 493 (2007).

\bibitem{Baier:1996sk}
  R.~Baier, Y.~L.~Dokshitzer, A.~H.~Mueller, S.~Peigne and D.~Schiff, Nucl.\ Phys.\  B {\bf 484}, 265 (1997).

\bibitem{Wiedemann:2000za}
  U.~A.~Wiedemann, Nucl.\ Phys.\  B {\bf 588}, 303 (2000).

\bibitem{Qiu:1990xa}
  J.~w.~Qiu and G.~F.~Sterman, Nucl.\ Phys.\  B {\bf 353}, 105 (1991).

\bibitem{Arnold:2001ba}
  P.~B.~Arnold, G.~D.~Moore and L.~G.~Yaffe, JHEP {\bf 0111}, 057 (2001).

\bibitem{Adams:2003kv} 
  J.~Adams {\it et al.} [STAR Collaboration], Phys.\ Rev.\ Lett.\  {\bf 91}, 172302 (2003).

\bibitem{Adler:2006bv} 
  S.~S.~Adler {\it et al.} [PHENIX Collaboration], Phys.\ Rev.\ C {\bf 75}, 024909 (2007).

\bibitem{Vitev:2002pf} 
  I.~Vitev and M.~Gyulassy, Phys.\ Rev.\ Lett.\  {\bf 89}, 252301 (2002).

\bibitem{Gyulassy:2004zy}
  M.~Gyulassy and L.~McLerran, Nucl.\ Phys.\  A {\bf 750}, 30 (2005).

\bibitem{Shuryak:2003xe} 
  E.~Shuryak, Prog.\ Part.\ Nucl.\ Phys.\  {\bf 53}, 273 (2004).

\bibitem{Romatschke:2007mq} 
  P.~Romatschke and U.~Romatschke, Phys.\ Rev.\ Lett.\  {\bf 99}, 172301 (2007).

\bibitem{MG}
  M.~Gyulassy, Talk given at the HCBM 2010 Conference, Budapest, Hungary, August 2010.

\bibitem{Maldacena:1997re}
  J.~M.~Maldacena, Adv.\ Theor.\ Math.\ Phys.\  {\bf 2}, 231 (1998) [Int.\ J.\ Theor.\ Phys.\  {\bf 38}, 1113 (1999)].

\bibitem{ches1}
  S.~S.~Gubser, D.~R.~Gulotta, S.~S.~Pufu and F.~D.~Rocha, JHEP {\bf 0810}, 052 (2008).

\bibitem{ches2}
  P.~M.~Chesler, K.~Jensen, A.~Karch and L.~G.~Yaffe, Phys.\ Rev.\  D {\bf 79}, 125015 (2009).

\bibitem{ches3}
  P.~M.~Chesler, K.~Jensen and A.~Karch,Phys.\ Rev.\  D {\bf 79}, 025021 (2009).

\bibitem{Gyulassy:2011zz} 
  M.~Gyulassy, A.~Buzzatti, A.~Ficnar, J.~Noronha and G.~Torrieri,
  EPJ Web Conf.\  {\bf 13}, 01001 (2011).

\bibitem{Adare:2010sp}
  A.~Adare {\it et al.}  [PHENIX Collaboration], Phys.\ Rev.\ Lett.\  {\bf 105}, 142301 (2010).

\bibitem{Horowitz:2011gd} 
  W.~A.~Horowitz and M.~Gyulassy, Nucl.\ Phys.\ A {\bf 872}, 265 (2011).

\bibitem{Betz:2011tu} 
  B.~Betz, M.~Gyulassy and G.~Torrieri, Phys.\ Rev.\ C {\bf 84}, 024913 (2011).

\bibitem{Pal:2012gf} 
  S.~Pal and M.~Bleicher, Phys.\ Lett.\ B {\bf 709}, 82 (2012).

\bibitem{Horowitz:2007ui} 
  W.~A.~Horowitz, J.\ Phys.\ G G {\bf 35}, 044025 (2008).

\bibitem{Betz:2012qq} 
  B.~Betz and M.~Gyulassy, Phys.\ Rev.\ C {\bf 86}, 024903 (2012).

\bibitem{Betz:2014cza} 
  B.~Betz and M.~Gyulassy, JHEP {\bf 1408}, 090 (2014) [JHEP {\bf 1410}, 043 (2014)].

\bibitem{Liu:2006he} 
H.~Liu, K.~Rajagopal and U.~A.~Wiedemann, JHEP {\bf 0703}, 066 (2007).

\bibitem{Baier:2006pt} 
R.~Baier, A.~H.~Mueller and D.~Schiff, Phys.\ Lett.\ B {\bf 649}, 147 (2007).

\bibitem{Renk:2005ta} 
T.~Renk and J.~Ruppert, Phys.\ Rev.\ C {\bf 72}, 044901 (2005).

\bibitem{Armesto:2004vz} 
N.~Armesto, C.~A.~Salgado and U.~A.~Wiedemann, Phys.\ Rev.\ C {\bf 72}, 064910 (2005).

\bibitem{Xu:2014tda} 
J.~Xu, J.~Liao and M.~Gyulassy, Chin.\ Phys.\ Lett.\  {\bf 32}, no. 9, 092501 (2015).

\bibitem{Buzzatti:2012dy} 
  A.~Buzzatti and M.~Gyulassy, Nucl.\ Phys.\ A {\bf 904-905}, 779c (2013).

\bibitem{Buzzatti:2012pe} 
  A.~Buzzatti and M.~Gyulassy, Nucl.\ Phys.\ A {\bf 910-911}, 490 (2013).

\bibitem{VISH2+1}
C.~Shen, U.~Heinz, P.~Huovinen and H.~Song, Phys.\ Rev.\ C {\bf 84}, 044903 (2011);
 Phys.\ Rev.\ C {\bf 82}, 054904 (2010).

\bibitem{Luzum:2008cw} 
  M.~Luzum and P.~Romatschke, Phys.\ Rev.\ C {\bf 78}, 034915 (2008) [Erratum-ibid.\ C {\bf 79}, 039903 (2009)].

\bibitem{BAMPS}
Z.~Xu and C.~Greiner, Phys.\ Rev.\ C {\bf 71}, 064901 (2005).

\bibitem{PHENIX_old}
 A.~Adare {\it et al.} (PHENIX Collaboration), Phys.\ Rev.\ Lett.\  {\bf 101}, 232301 (2008).

\bibitem{Collaboration:2011hf}
  J.~Jia (ATLAS Collaboration), J.\ Phys.\ G {\bf 38}, 124012 (2011).

\bibitem{CMSRAA}
  S.~Chatrchyan {\it et al.}  (CMS Collaboration),
  Eur.\ Phys.\ J.\ C {\bf 72}, 1945 (2012).

\bibitem{Ficnar:2011yj} 
A.~Ficnar, J.~Noronha and M.~Gyulassy, J.\ Phys.\ G {\bf 38}, 124176 (2011).

\bibitem{Finazzo:2014cna} 
S.~I.~Finazzo, R.~Rougemont, H.~Marrochio and J.~Noronha, 
JHEP {\bf 1502}, 051 (2015).

\bibitem{Xu}
J.~Xu, A.~Buzzatti and M.~Gyulassy, JHEP {\bf 1408}, 063 (2014).

\bibitem{Molnar}
D.~Molnar and D.~Sun, Nucl.\ Phys.\ A {\bf 910-911}, 486 (2013).

\bibitem{Betz:2015mlf} 
  B.~Betz, F.~Senzel, C.~Greiner and M.~Gyulassy, arXiv:1512.07443 [hep-ph].

\bibitem{BBMGvUSPhydro}
B.~Betz, J.~Noronha-Hostler, J.~Noronha, M.~Gyulassy, in preparation.

\bibitem{data2_RAA}
B.~Abelev {\it et al.}  [ALICE Collaboration], Phys.\ Lett.\ B {\bf 720}, 52 (2013).

\bibitem{data3_RAA}
S.~Chatrchyan {\it et al.}  [CMS Collaboration], Eur.\ Phys.\ J.\ C {\bf 72}, 1945 (2012).

\bibitem{Aad:2013xma} 
  G.~Aad {\it et al.} [ATLAS Collaboration], JHEP {\bf 1311}, 183 (2013).

\bibitem{Noronha-Hostler:2016eow} 
  J.~Noronha-Hostler, B.~Betz, J.~Noronha and M.~Gyulassy, arXiv:1602.03788 [nucl-th].

\bibitem{Noronha-Hostler:2013gga} 
  J.~Noronha-Hostler, G.~S.~Denicol, J.~Noronha, R.~P.~G.~Andrade and F.~Grassi,
  Phys.\ Rev.\ C {\bf 88}, 044916 (2013).

\bibitem{Noronha-Hostler:2014dqa} 
  J.~Noronha-Hostler, J.~Noronha and F.~Grassi, Phys.\ Rev.\ C {\bf 90}, no. 3, 034907 (2014).

\bibitem{Aad:2015lwa} 
  G.~Aad {\it et al.} [ATLAS Collaboration], Phys.\ Rev.\ C {\bf 92}, no. 3, 034903 (2015).

\bibitem{Aamodt:2011by} 
  K.~Aamodt {\it et al.} [ALICE Collaboration], Phys.\ Lett.\ B {\bf 708}, 249 (2012).

\bibitem{data2_v2}
B.~Abelev {\it et al.}  [ALICE Collaboration], Phys.\ Lett.\ B {\bf 719}, 18 (2013).

\bibitem{data3_v2}
S.~Chatrchyan {\it et al.}  [CMS Collaboration], Phys.\ Rev.\ Lett.\  {\bf 109}, 022301 (2012).

\bibitem{data4}
G.~Aad {\it et al.}  [ATLAS Collaboration], Phys.\ Lett.\ B {\bf 707}, 330 (2012).


\bibitem{Niemi:2012aj} 
  H.~Niemi, G.~S.~Denicol, H.~Holopainen and P.~Huovinen,
  Phys.\ Rev.\ C {\bf 87}, no. 5, 054901 (2013).

\bibitem{Retinskaya:2013gca} 
  E.~Retinskaya, M.~Luzum and J.~Y.~Ollitrault, Phys.\ Rev.\ C {\bf 89}, no. 1, 014902 (2014).

\end{thebibliography}
%

\end{document}